\documentclass{iau}
\usepackage{graphicx}
\usepackage{subcaption}

\title[Planetary Systems in Star Clusters] 
{Planetary Systems in Star Clusters}

\author[Maxwell Xu Cai, Rainer Spurzem \& M.B.N. Kouwenhoven]   
{Maxwell Xu Cai$^{1,2}$, Rainer Spurzem$^{1,3}$
 \and M.B.N. Kouwenhoven$^2$}

\affiliation{$^1$National Astronomical Observatories, Chinese Academy of Sciences, \\ Beijing 100012,
20A Datun Road, Chaoyang District, Beijing, P.R. China \\ email: {\tt maxwell@nao.cas.cn} \\[\affilskip]
$^2$Kavli Institute for Astronomy and Astrophysics, Peking University \\ 5 Yi He Yuan Road, Haidian District, Beijing \\[\affilskip]
$^3$Astronomisches Rechen-Institut, Zentrum f\"ur Astronomie, University of Heidelberg, M\"onchhofstrasse 12-14, 69120 Heidelberg, Germany }

\pubyear{2014}
\volume{312}  
\pagerange{119--126}
\jname{Star Clusters and Black Holes in Galaxies Across Cosmic Time}
\editors{A.C. Editor, B.D. Editor \& C.E. Editor, eds.}
\begin{document}

\maketitle

\begin{abstract}
In the solar neighborhood, where the typical relaxation timescale is larger than the cosmic age, at 
least 10\% to 15\% of Sun-like stars have planetary systems with Jupiter-mass planets. In contrast, 
dense star clusters, charactered by frequent close encounters, have been found to host very few 
planets. We carry out numerical simulations with different initial conditions to investigate the 
dynamical stability of planetary systems in star cluster environments.
\end{abstract}

\firstsection 
\section{Introduction}
Recent studies suggest that most stars are formed in groups and clusters 
(\cite[Lada \& Lada 2003]{lada2003}). The evolution of star clusters is driven 
by internal (e.g. two-body relaxationi, mass loss by stellar evolution),
as well as external processes (e.g. tidal forces of a galaxy)
(\cite[Spitzer 1987]{spitzer87}, \cite[Elson et al. 1987]{elson87}). Since these mechanisms
finally lead to the 
dissolution of clusters they have a 
finite life time. So, if our sun has been born in an open star cluster, the solar planetary system
may also have formed before the dissolution of its parent star cluster.

Observations of exoplanets have been possible since the early 1990s.
As of December 2014, more than 1,800 exoplanets have been discovered, 
among which are 473 multi-planet systems like our own solar system ({\tt http://exoplanet.eu}). 
Nevertheless, most exoplanets are found in the solar neighborhood; very few exoplanets 
(e.g. Kepler-66, Kepler-67) are found in star clusters. Star clusters, especially cores of globular clusters, 
are characterized by high density of stars and high velocity dispersions, which results in frequent 
close encounters. Current observations suggest that formation of planetary systems in dense
regions of star clusters. This could be due to instability against external gravitational
perturbations, though there may also be some instrumental biases due to detection limits.

Previous studies include direct $N$-body simulation (\cite[Spurzem et al. 2009]{spurzem09}) and 
Monte-Carlo simulations (\cite[Hao et al. 2013]{hao13}). 
We present numerical simulations
to investigate the dynamical stability of planetary systems in star 
clusters. Coupling planetary system with star cluster dynamics is a multi-scale problem, due to
time scales ranging from $\sim 1-100$ for planetary orbits to $\sim 10^4 - 10^6$ years for stars
orbiting in the cluster. Full direct coupled simulations are expensive, though not impossible
(see e.g. \cite[Spurzem et al. 2009]{spurzem09}); but frequently Monte Carlo approaches are used
(see e.g. \cite[Hao et al. 2013]{hao13}). The {\tt AMUSE} framework is a convenient toolbox to study
such a problem
(\cite[Portegies Zwart et al., 2013]{portegies13}; 
\cite[Portegies Zwart et al., 2009]{portegies09}). Within {\tt AMUSE} we integrate star cluster dynamics 
directly with {\tt NBODY6++} (\cite[Spurzem 1999]{spurzem99}), while the planets are 
integrated with {\tt MERCURY6} (\cite[Chambers 1999]{chambers99}). The perturbations are 
implemented with the {\tt BRIDGE} scheme (\cite[Fujii et al., 2007]{fujii07}; 
\cite[Pelupessy et al,. 2013]{Pelupessy13}). All coupling is done within {\tt AMUSE} framework.

\section{Results and Conclusions}
For the initial setup of our system we follow the previously cited papers (Plummer sphere for
stellar density, $N=2\mathrm{k}$ and $N=10\mathrm{k}$ stars, virial radius $R_v = 1$ pc; 
(A) current gas giant configuration or (B) equal mass and mutual Hill radii separation 
for the planets). For case (B) the inner planet has $M=1 M_J$ and $a=1$ AU.
We distribute 50 (100) planetary systems randomly across the 
$N=2\mathrm{k}$ ($N=10\mathrm{k}$) cluster, to sample different positions and environments of
planetary systems in star clusters.

\begin{figure}
  \centering
  \begin{subfigure}[b]{0.32\textwidth}
    \centering
    \includegraphics[width=\textwidth]{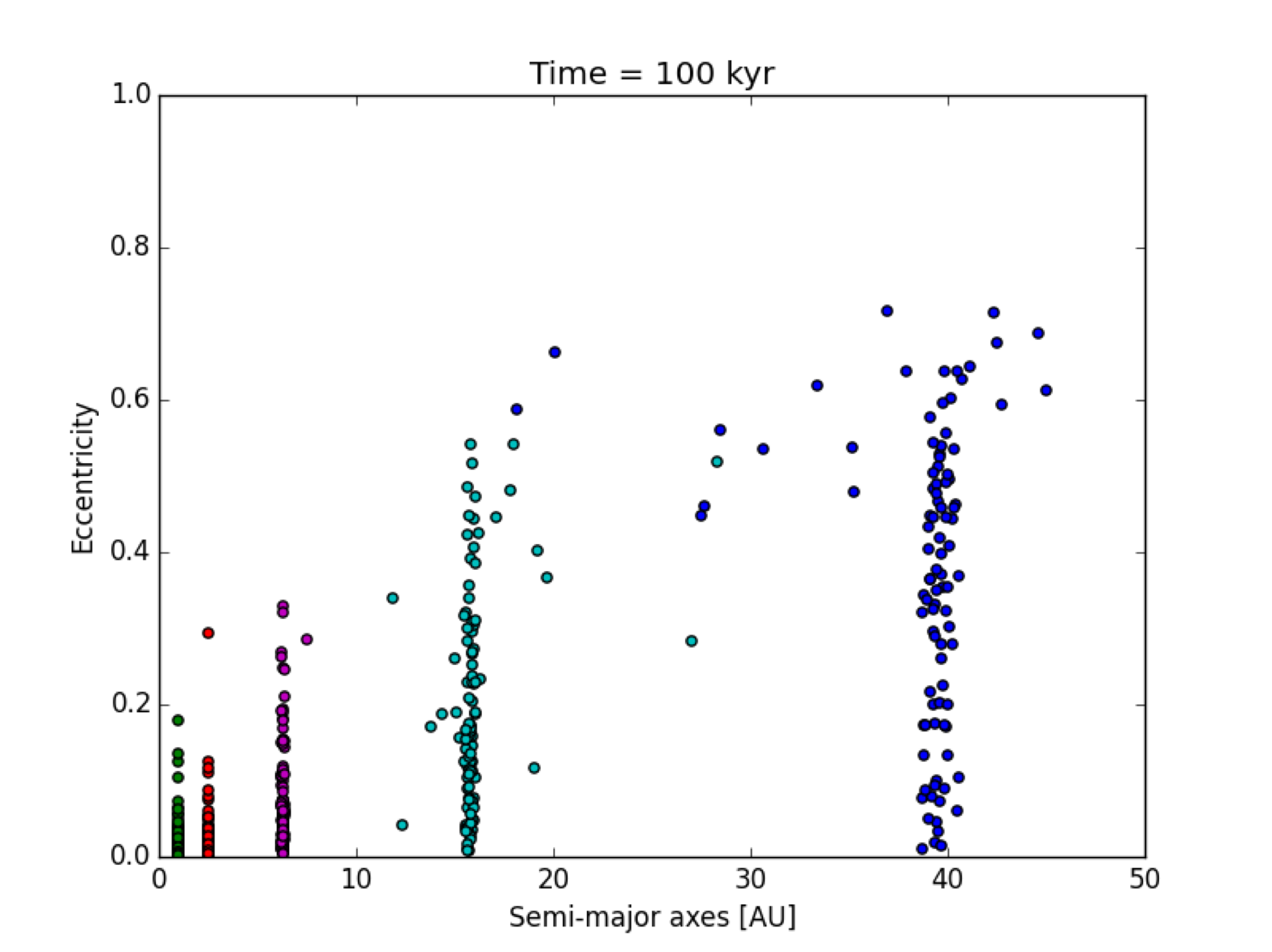}
    \caption{}
    \label{fig:aeplot}
  \end{subfigure}
  \begin{subfigure}[b]{0.32\textwidth}
    \centering
    \includegraphics[width=\textwidth]{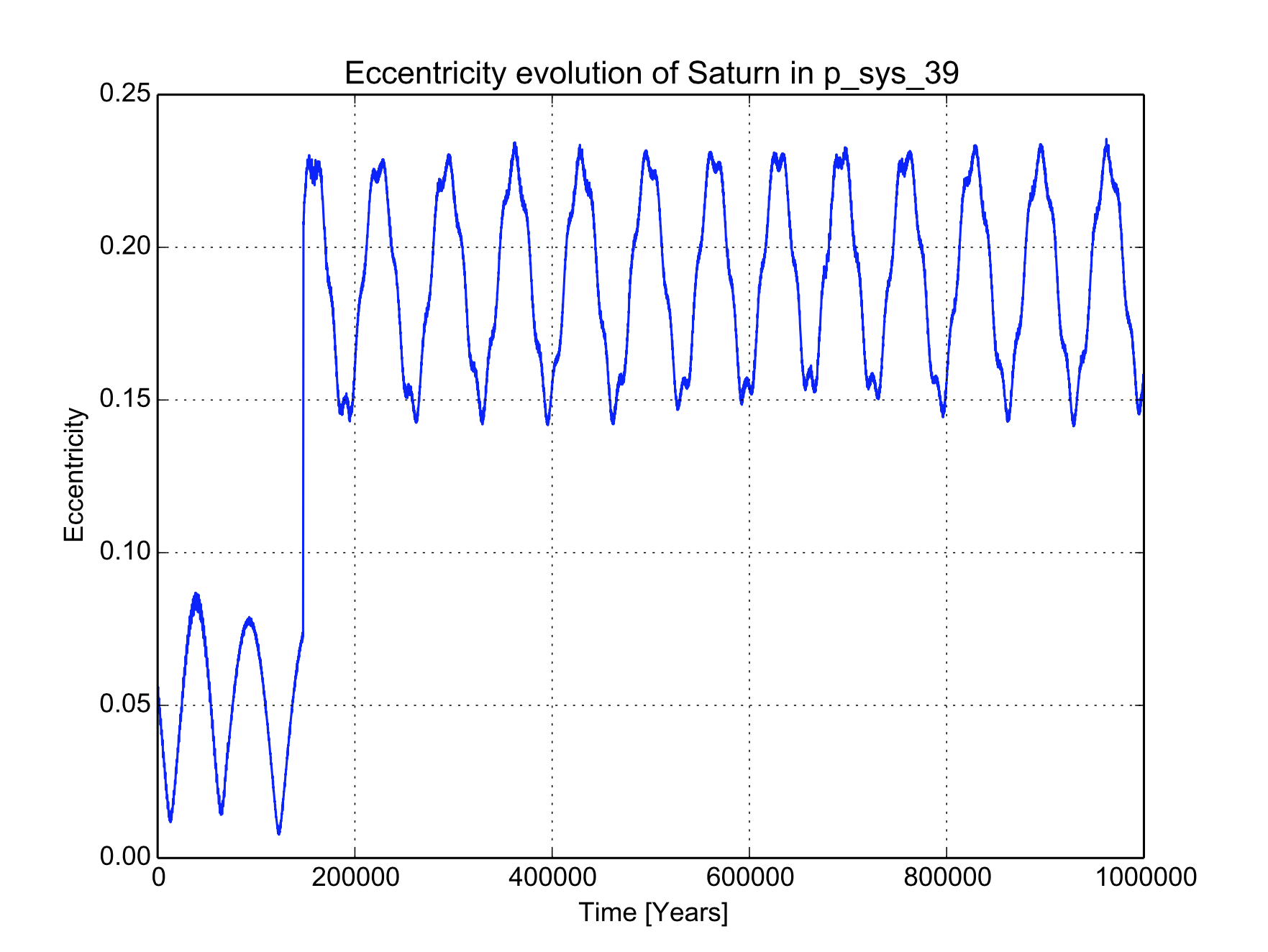}
    \caption{}
    \label{fig:ecc}
  \end{subfigure}
  \begin{subfigure}[b]{0.32\textwidth}
    \centering
    \includegraphics[width=\textwidth]{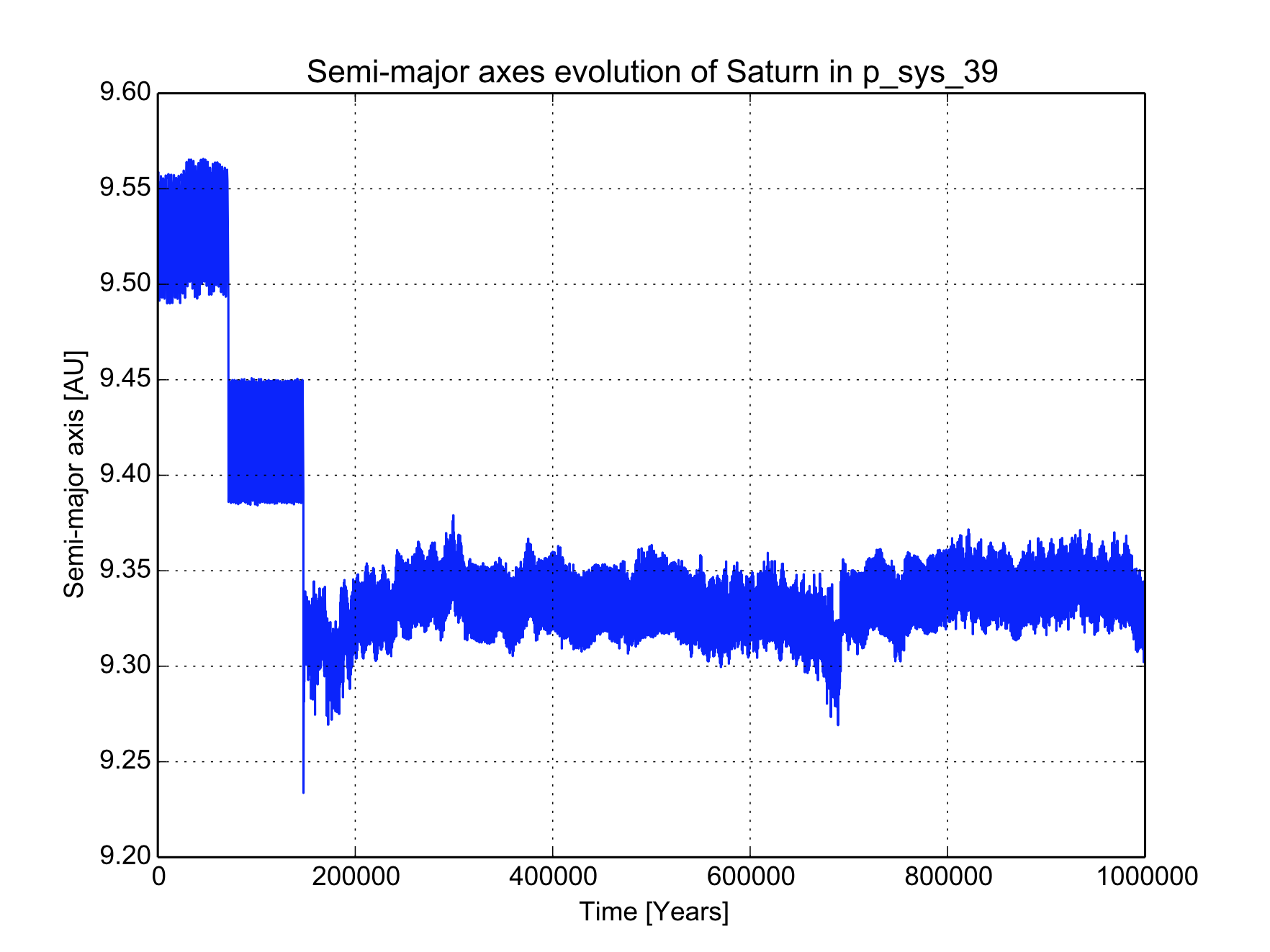}
    \caption{}
    \label{fig:semi}
  \end{subfigure}
  \caption{(a) Diffusion of the $a-e$ space at $t=100$ kyr of a 
simulation; (b) Eccentricity evolution of Saturn in one of the planetary systems; (c) Corresponding 
semi-major axis evolution of Saturn in the same planetary system as (b).}\label{fig:results}
\end{figure}

Figure~\ref{fig:aeplot} shows the $a-e$ space (semi-major axis verses 
eccentricity) at the end of the $N=10\mathrm{k}$ simulation (corresponding to $T = 0.1$ Myr). 
The evolution of semi-major axes and eccentricities of Saturn in the $N=2\mathrm{k}$ 
simulation are shown in Figures~\ref{fig:ecc} and~\ref{fig:semi}, respectively.
Our simulations show that generally
planetary systems have difficulty surviving within dense cluster environments, but some compact
ones are sheltered in the potential well of their hosts from disruption. 

We acknowledge support by NAOC CAS through the Silk Road Project and (RS) through the Chinese 
Academy of Sciences Visiting Professorship for Senior International Scientists, 
Grant Number 2009S1-5. We are grateful for the supports from the {\tt AMUSE} team, especially 
Simon Portegies Zwart, Inti Pelupessy, Arjen van Elteren and Nathan de Vries for various 
useful discussions. MXC and MBNK acknowledge the {\tt AMUSE} team for supporting their visits to Leiden, and Michiko Fujii for useful discussions.

\end{document}